\documentclass{ws-procs11x85}

\usepackage{ws-procs-thm}           
\usepackage{booktabs}
\usepackage{hyperref}
\let\captionbox\undefined
\usepackage{subcaption}
\begin{document}

\title{CloudPred: Predicting Patient Phenotypes From Single-cell RNA-seq}

\author{Bryan He$^1$, Matthew Thomson$^{2,3}$, Meena Subramaniam$^4$,\\Richard Perez$^4$, Chun~Jimmie~Ye$^4$, James Zou$^{1,5,6,7}$}

\address{
  $^1$Department of Computer Science, Stanford University \\
  $^2$Division of Biology and Biological Engineering, California Institute of Technology \\
  $^3$Beckman Center for Single-cell Profiling and Engineering \\
  $^4$Department of Epidemiology and Biostatistics, University of California, San Francisco \\
  $^5$Department of Electrical Engineering, Stanford University \\
  $^6$Department of Biomedical Data Science, Stanford University \\
  $^7$Chan-Zuckerberg Biohub \\
  Email: bryanhe@stanford.edu, jamesz@stanford.edu
}

\begin{abstract}
Single-cell RNA sequencing (scRNA-seq) has the potential to provide powerful, high-resolution signatures to inform disease prognosis and precision medicine.
This paper takes an important first step towards this goal by developing an interpretable machine learning algorithm, CloudPred, to predict individuals' disease phenotypes from their scRNA-seq data.
Predicting phenotype from scRNA-seq is challenging for standard machine learning methods---the number of cells measured can vary by orders of magnitude across individuals and the cell populations are also highly heterogeneous.
Typical analysis creates pseudo-bulk samples which are biased toward prior annotations and also lose the single cell resolution.
CloudPred addresses these challenges via a novel end-to-end differentiable learning algorithm which is coupled with a biologically informed mixture of cell types model.
CloudPred automatically infers the cell subpopulation that are salient for the phenotype without prior annotations.
We developed a systematic simulation platform to evaluate the performance of CloudPred and several alternative methods we propose, and find that CloudPred outperforms the alternative methods across several settings.
We further validated CloudPred on a real scRNA-seq dataset of 142 lupus patients and controls.
CloudPred achieves AUROC of 0.98 while identifying a specific subpopulation of CD4 T cells whose presence is highly indicative of lupus.
CloudPred is a powerful new framework to predict clinical phenotypes from scRNA-seq data and to identify relevant cells.   
\end{abstract}


\copyrightinfo{\copyright\ 2021 The Authors. Open Access chapter published by World Scientific Publishing Company and distributed under the terms of the Creative Commons Attribution Non-Commercial (CC BY-NC) 4.0 License.}

\section{Introduction}
Single-cell RNA-seq (scRNA-seq) is a powerful approach for profiling the composition and expression profiles of heterogeneous cell populations and has led to many insights in molecular and cellular biology  \cite{tang2011nm,eberwine2014promise}.
In addition to fundamental biology, scRNA-seq has the potential to become a valuable tool for precision health by revealing how specific cells correlate with symptoms and by informing diagnosis and treatment.
To realize the tremendous clinical utility of scRNA-seq, a key step is to be able to directly predict clinical phenotypes, such as disease status, from patients' scRNA-seq data. 

Prediction from scRNA-seq data is challenging and is not well-addressed by current computational biology methods for several reasons.
First, the data collected from each patient corresponds to the expression profile of a variable number of cells---some patients might only have hundreds of cells sequenced while others could have thousands of cells.
Second, the number of patients available is typically small---often dozens---and the expression profile of each cell is noisy and high-dimensional.
Standard prediction algorithms generally require that each sample/patient has the same number of measurements and cannot be directly applied to scRNA-seq data \cite{zou2019primer}.
An alternative approach is to average across the single cells to generate a pseudo-bulk expression for each patient; however this loses important information of cell-cell variation \cite{luecken2019current}.
Partially due to these challenges, the standard analysis of scRNA-seq data is primarily limited to unsupervised learning (e.g. clustering, trajectory inference) or simpler differential expression analysis \cite{luecken2019current,abid2018exploring}.
To the best of our knowledge, there has not been a published methodology that can directly predict human patient phenotypes from scRNA-seq.   

\begin{figure}[thbp!]
    \centering
    \includegraphics[scale=0.83]{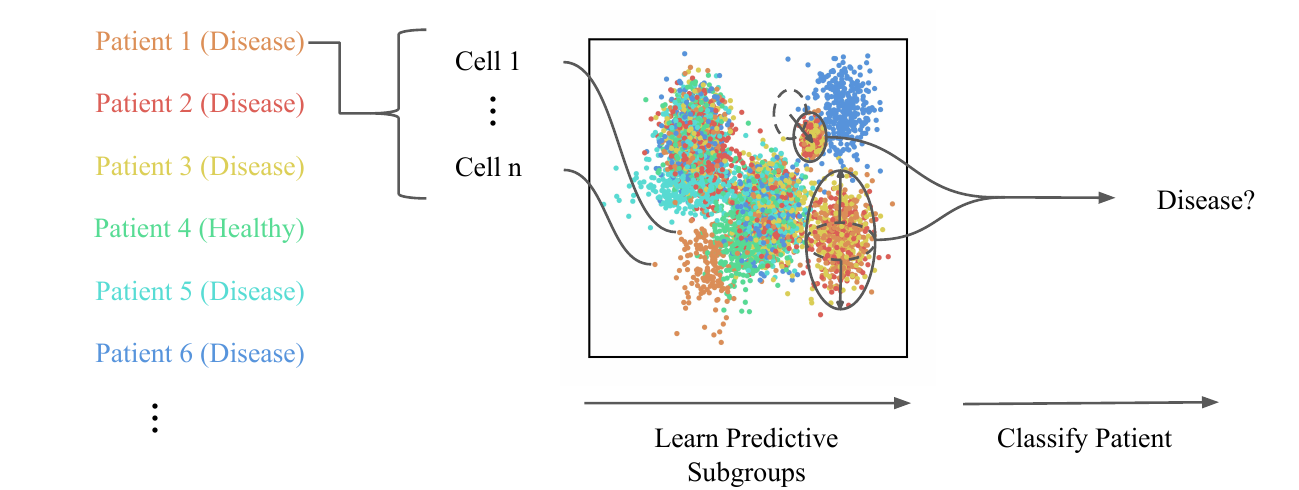}
    \caption{Schematic of the CloudPred algorithm. The data from each patient is a cloud of variable number of points/cells (different patients represented by different colors here). CloudPred automatically identifies subpopulation of cells---highlighted by the ellipses---whose variation is reliably predictive of the phenotype of the patient, e.g. disease status. After training on labeled patients, CloudPred can then reliably predict phenotypes on new patients.
}
    \label{fig: pipeline}
\end{figure}

In this work, we introduce CloudPred, a new algorithm to predict phenotype from scRNA-seq data (Fig.~\ref{fig: pipeline}). CloudPred is designed to automatically address the challenge that different samples have different number of points. CloudPred also identifies variations in specific sub-populations of cells which could be prognostic of disease. The name CloudPred arises from the observation that the scRNA-seq data for a sample can be thought of as a point cloud: a single point in the point cloud represents the expression profile of a cell, and all cells sampled for the sample make up the point cloud. CloudPred leverages this geometric representation in its algorithm, and makes reliable predictions even when trained on only dozens of samples.
\section{Methods}
\subsection{Overview}
CloudPred automatically identifies  subpopulations of cells that are predictive of the relevant phenotype (Fig.~\ref{fig: pipeline}). It takes as input a collection of scRNA-seq data from multiple patients and splits the data into train, validation and test patients. The goal is to train a classifier that can predict the patient's phenotype (e.g. disease or healthy) from his/her scRNA-seq. The data from each patient has the scRNA expression profile of a variable number of cells, which could differ by orders of magnitude across patients. CloudPred models each patient's data as a mixture of Gaussians.
The mean and covariance of each mixture component characterize one candidate subpopulation of cells. From a given set of means and covariances, CloudPred estimates the abundance of each cell population in one patient. These abundances forms a set of patient specific features and is used to predict the patient phenotype.
These abundances can be thought of as a continuous bag-of-features, which have been used in image-based models \cite{lazebnik2008supervised,chidester2018discriminative}.

Standard scRNA-seq analysis often identifies cell populations using prior knowledge and marker genes. Similarly, in standard mixture model, the components (e.g. mean and covariance) of the mixtures is fixed after fitted on the data. In contrast, CloudPred automatically learns the populations whose variation across patients are the most predictive of the phenotype.  
All of the model parameters---the means and covariances of the Gaussians---are
 trained end-to-end using stochastic gradient descent to minimize prediction error on the training data. This training procedure allows the algorithm to identify the phenotype-relevant cell populations in an unbiased data driven manner.

\subsection{CloudPred Prediction Pipeline}
We now describe how CloudPred makes a prediction on a patient, assuming that the parameters have already been learned, and we describe the procedure for initializing and learning the parameters in the following section.
We represent a single patient as a multiset $\{\mathbf{x}_i \in \mathbb{R}^d\}_{i=1}^{n}$, where $n$ is the number of cells/points from the patient, and $d$ is the dimension of the feature space, which for scRNA-seq is the number of measured genes. 

CloudPred begins by modeling the individual points as samples from a mixture of Gaussians \cite{zhang2019valid}.
This is represented as $m$ subpopulations with mean $\mu_j$, covariance $\Sigma_j$, and weight $\mathbf{w}_j$, which are shared across patients.
To reduce the number of parameters to learn, we restrict the covariance matrices to diagonal matrices.
The probability that subpopulation $j$ generates a point $\mathbf{x}$ is then
\begin{align}
  \Pr(\mathbf{x}\mid \mu_j,\Sigma_j) \sim \mathcal{N}(\mu_j, \Sigma_j).
\end{align}

CloudPred uses this model to probabilistically assign points to clusters.
The probability that a point $\mathbf{x}_i$ is assigned to cluster $j$ is given by
\begin{align}
  \mathbf{p}_{ij} = \frac{\mathbf{w}_j\Pr(\mathbf{x}\mid \mu_j,\Sigma_j)}{\sum_{k=1}^{m}\mathbf{w}_k\Pr(\mathbf{x}\mid \mu_k,\Sigma_k)}.
\end{align}

The estimated prevalence of the subpopulations is then
\begin{align}
    \mathbf{s} = \frac{\sum_{i=1}^{n} \mathbf{p}_i}{n}
\end{align}
where $\mathbf{p}_i$ is the $m$ dimensional vector of the $\mathbf{p}_{ij}$ giving the probability that cell $i$ belongs to each cluster, and $\mathbf{s}$ is the $m$ dimensional vector where $s_j$ is the total estimated number of cells in cluster $j$. Recall that $m$ denotes the number of clusters, and it can be specified using any standard single cell clustering methodology.  

Notice that $\mathbf{s}$ is now of fixed dimension, which makes it appropriate for use in standard machine learning techniques.
As a result, we can apply a classifier $f_\theta$ to make a prediction on the outcome.
In our experiments, we use a  classifier with a quadratic term:
\begin{align}
  f_{\theta}(\mathbf{s}) = f_{a,b,c}(\mathbf{s}) = \sum_{j=1}^{m}a_js_j + \sum_{j=1}^{m}b_js_j^2 + c,
\end{align}
where $\theta = \{a_j, b_j, c\}$ are the  parameters of the classifier. The quadratic term in $f_{\theta}$ gives the model additional flexibility to capture scenarios where having too few or too many cells of one cluster could be indicative of the phenotype. 
We use cross entropy loss between the classifiers prediction and the true label, given by
\begin{align}
-y\log(p) + (1 - y)\log(1 - p),
\end{align}
where $y\in\{0, 1\}$ is the true label, and $p=1 / \left(1 + e^{f_\theta(\mathbf{s})}\right)$ is the classifier's predicted probability that the example is a positive example.

\subsection{Training Procedure}

We begin by selecting an initialization for the parameters.
We use the EM algorithm to compute an initial estimate of the centers, covariances, and weights for each of the mixture components, providing a set of candidate subpopulations \cite{scikit-learn}.
The classifier parameters $\theta = \{a_i, b_i, c\}$ are initialized randomly.
Notice that this initialization is unsupervised; the outcomes of the patients have not yet been considered.

We then train all parameters of the model end-to-end using stochastic gradient descent with a learning rate of $10^{-4}$ for 1,000 epochs.
During training, we use the cross entropy loss between CloudPred's prediction and the true class, and we additionally track the loss on the validation set.
The final model selected is the model with the lowest cross-entropy loss on the validation set.
The entire prediction pipeline is differentiable, so we use the automatic differentiation in the Pytorch library in Python to compute the gradients \cite{paszke2017automatic}.
During the training process, the parameters in the logistic regression model, $\mathbf{a}$, $\mathbf{b}$, and $c$, along with the mixture model parameters, $\mu$, $\Sigma$, and $\mathbf{w}$, are updated.

\subsection{Alternative Prediction Methods}
Because prediction using scRNA-seq is relatively unexplored, we propose several alternative methods to evaluate alongside CloudPred.
These alternatives also serve as ablations that enable us to assess the importance of different modeling choices in CloudPred.

The first alternative method (\emph{independent}) treats cells from a patient as independent, makes a prediction for each cell separately, and then averages the cell-wise predictions.
The predictions for this model are given by $\frac{1}{n}\sum_{i=1}^{n}\mathbf{w}^T\mathbf{x}_i$, where $\mathbf{w}$ are the parameters of the model.

Next, we propose a method that fits two Gaussian mixture models, one for all of the diseased patients and one for all of the healthy controls. This models the canonical cell populations for the two different phenotypes. For a new patient, we compute the likelihood of its scRNA-seq data under each mixture model and select the phenotype with higher likelihood as the prediction.  
This approach is motivated by the analysis in Chen et al.\cite{chen2018biorxiv} which focuses on identifying disease signatures in RNA-seq data; we denote this method \emph{Mixture (class)}.
We separately compare with a more fine-grained mixture modeling method, where a Gaussian mixture is fit to each patient separately, rather than to all of the disease (or healthy) patients as in \emph{Mixture (class)}. 
Then, for a new patient, we compute the likelihood of its scRNA-seq data under each of the patient mixtures, and select the phenotype with the most similar training patients (i.e. higher total likelihood); we denote this method \emph{Mixture (patient)}.
The key difference between CloudPred and these two mixture models is that CloudPred learns the population component simultaneously while making the predictions, whereas the more standard mixture models rely on unsupervised clustering to fix the population components. 

Finally, recent works in deep learning have introduced Deep Sets, which represents the most general family of permutation-invariant functions, to make predictions on sets. \cite{zaheer2017nips,qi2017cvpr}
In this setting, each set corresponds to scRNA-seq from one patient.
While the family of functions Deep Sets would eventually be able to learn an accurate model on a sufficiently large training set, it is unable to generalize well without many training examples.
The trained classifier also suffers from the lack of interpretability typical for deep learning models \cite{ghorbani2019interpretation}. 

\subsection{Experimental protocol}

To ensure that CloudPred, along with the alternative methods generalize to new patients, all experiments hold out a validation and test set.
Only the training set is used to learn model parameters, and the validation set is used to determine hyperparameters.
The validation set is used to select the number of clusters in the mixture model for CloudPred, the generative model by class, and the generative model by patient, and the number of hidden states in the DeepSets implementation; the linear model does not have a notion of clusters, so no hyperparameters are selected for it.
This hyperparameters is selected from 5, 10, 15, 20, and 25.
The validation set is also used to allow early stopping for the methods trained with stochastic gradient descent (CloudPred, independent, and Deep Sets). 
\begin{figure}[htbp]
    \begin{center}
        \begin{subfigure}[b]{0.35\textwidth}
            \begin{center}
                $\underbrace{\includegraphics[scale=0.4]{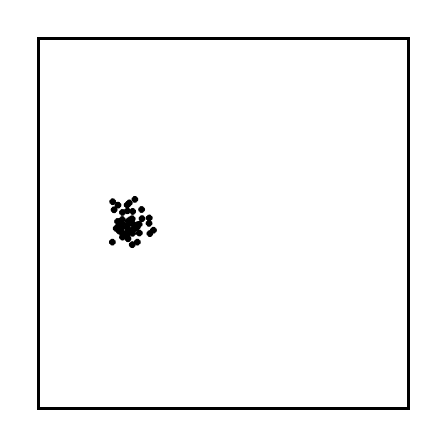}}_{\textrm{Normal}}$
                $\underbrace{\includegraphics[scale=0.4]{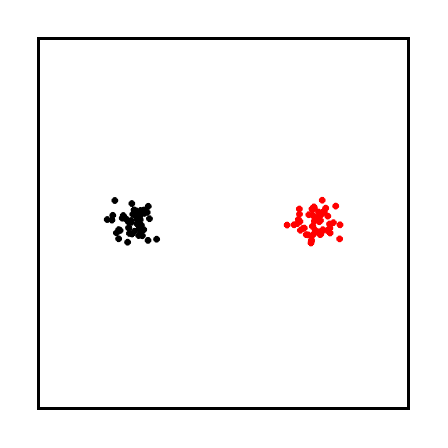}}_{\textrm{Disease}}$
                \caption{Basic case}
            \end{center}
        \end{subfigure}
        \begin{subfigure}[b]{0.60\textwidth}
            \begin{center}
                $\underbrace{\includegraphics[scale=0.4]{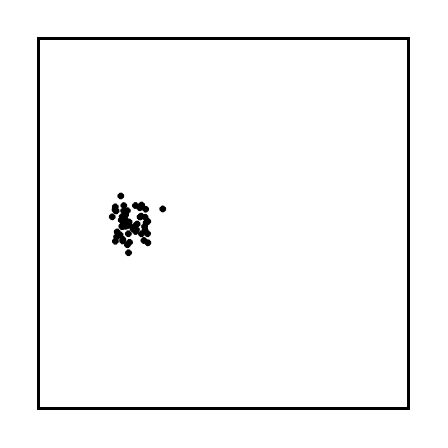}
                             \includegraphics[scale=0.4]{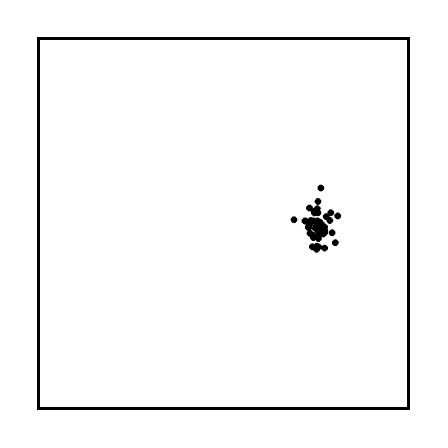}}_{\textrm{Normal}}$
                $\underbrace{\includegraphics[scale=0.4]{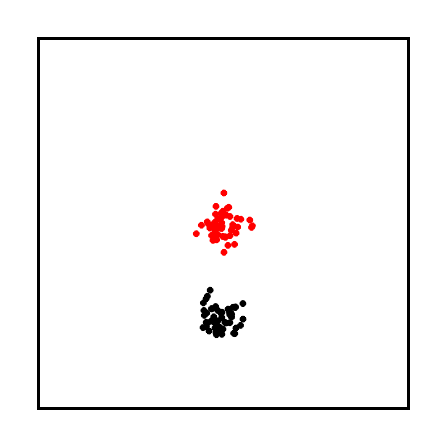}
                             \includegraphics[scale=0.4]{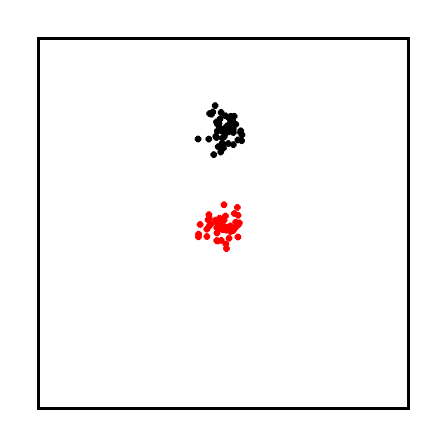}}_{\textrm{Disease}}$
                \caption{Patient-to-patient variation}
                \label{fig: example_variation}
            \end{center}
        \end{subfigure} \\
        
        \begin{subfigure}[b]{0.35\textwidth}
            \begin{center}
                $\underbrace{\includegraphics[scale=0.4]{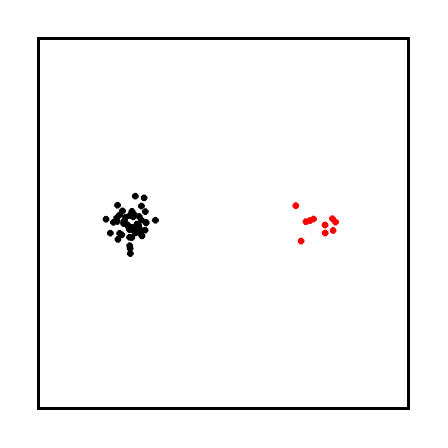}}_{\textrm{Normal}}$
                $\underbrace{\includegraphics[scale=0.4]{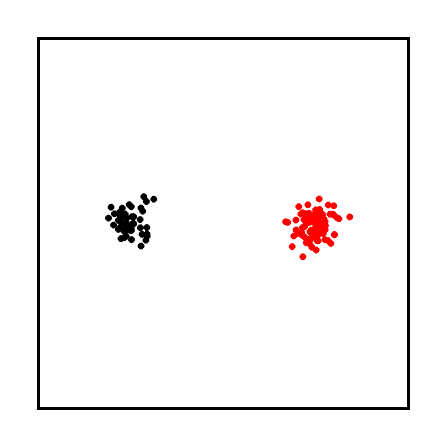}
                             \includegraphics[scale=0.4]{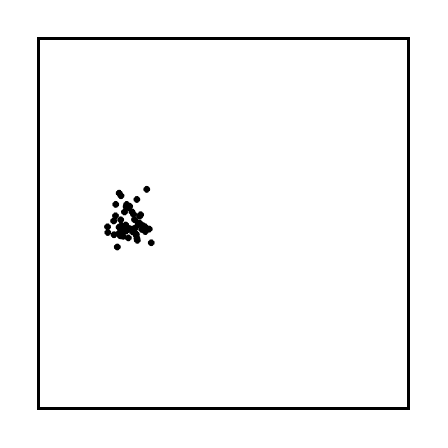}}_{\textrm{Disease}}$
                \caption{Multiple signatures}
                \label{fig: example_multiple}
            \end{center}
        \end{subfigure}
        \begin{subfigure}[b]{0.60\textwidth}
            \begin{center}
                $\underbrace{\includegraphics[scale=0.4]{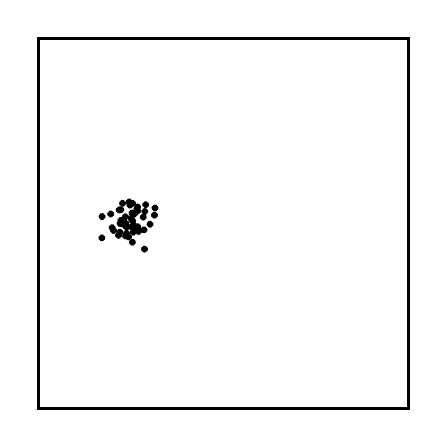}
                             \includegraphics[scale=0.4]{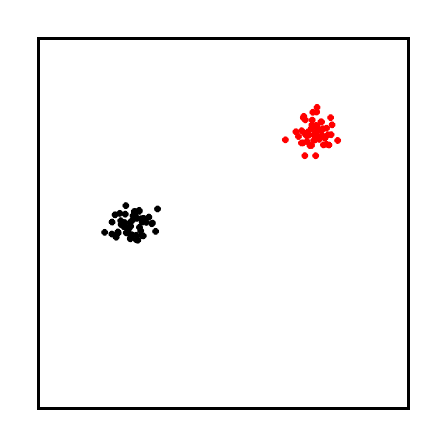}
                             \includegraphics[scale=0.4]{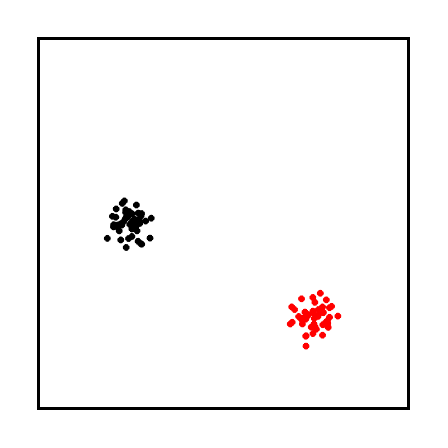}}_{\textrm{Normal}}$
                $\underbrace{\includegraphics[scale=0.4]{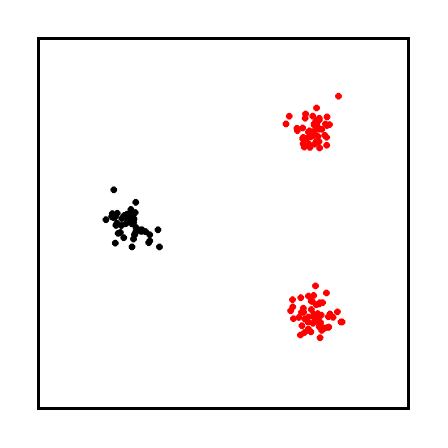}}_{\textrm{Disease}}$
                \caption{Interaction between subpopulations}
                \label{fig: example_interaction}
            \end{center}
        \end{subfigure}
        \caption{Examples of how variation in the point cloud can indicate disease status or another phenotype of interest. Cellular subpopulations of interest are highlighted in red. Each square illustrates the scRNA-seq of one example patient. (a) The presence of an additional subpopulation of cells is correlated with disease. (b) There may be patient-to-patient variation, where variations in many subpopulations of cells (black dots) are not relevant to the disease status. (c) The disease may have multiple signatures; for example, both an increase and an absence of a subpopulation may indicate disease. (d) There may be interactions between cell populations; for example, a disease may require multiple subpopulations of cells to be present.}
        \label{fig: examples}
    \end{center}
\end{figure}
\vspace{1.0cm}
\section{Results}

\subsection{Benchmarking performance with simulation study}

\paragraph{Interactions between cell populations complicate prediction}
We first identify several settings to conceptually illustrate the complexities that can arise in predicting from scRNA-seq point clouds.
In Fig.~\ref{fig: examples}, we show subpopulations relevant to the target of interest in red; this coloring is only for visualization and would not be available for making predictions.
A simple setting is shown in Fig.~\ref{fig: examples}a, where there is a subpopulation whose presence is indicative of disease.
Many methods, including the baselines we propose and evaluate here, are able to handle this simple scenario.
However, more complex interactions are possible in point clouds.
First, there may be patient-to-patient variations, shown in Fig.~\ref{fig: examples}b, where many subpopulations are not relevant to the target of interest.
Second, the disease may have multiple signatures, shown in Fig.~\ref{fig: examples}c. In our example, either an increase or a lack of a certain subpopulation can indicate disease.
Third, there may be interactions between subpopulations, where multiple subpopulations need to be present to indicate disease as in Fig.~\ref{fig: examples}d.
Finally, we note that these complications can occur together, further increasing the difficulty of identifying relevant subpopulations. The four scenarios that we highlight here crystallize the complexities that has been observed in scRNA-seq analysis and has been discussed in literature \cite{brodin2017nri,brodin2015cell,shen2016cell}. 

\paragraph{Simulation Setup}
Because scRNA-seq data from patients is still limited and the ground truth disease causing mechanism is often unclear, we first develop a systematic simulation framework to evaluate the prediction methods. Our simulations are motivated by the possible interaction scenarios described in Fig.~\ref{fig: examples}.
To simulate healthy and diseased patients, we generate semi-synthetic data using real scRNA-seqs from 94,655 cells consisting of 10 different cell types isolated from cell sorting---B cells, CD14 monocytes, CD34, CD4 helper T, CD56 NK, Cytotoxic T, Memory T, Naive cytotoxic, Naive T, Regulatory T from Chen et al. \cite{chen2018biorxiv}

Each synthetic patient is generated by taking a random sample of the scRNA-seq of actual cells from these 10 groups while adding additional noise. 
Across different simulations, we vary the amount of patient-to-patient variation, vary the prevalence of two disease signatures, and simulate the presence of interactions between cell types.
In the simulations, we use training sets ranging from 10 to 250 patients to learn parameters, a validation set of 100 patients is used to select hyperparameters, and evaluate on a test set of 100 patients.
For each of the simulations, we report performance as the area under the receiver operating characteristics (AUROC).
All experiments are run for 25 random splits of the dataset, and uncertainties represent 95\% confidence intervals computed with 10,000 bootstrapped samples \cite{efron1994introduction}.

\paragraph{Simulation Results}

\begin{figure}[htbp]
    \begin{center}
        \begin{subfigure}[b]{0.35\textwidth}
            \begin{center}
                \hspace{0.5cm}
                \raisebox{0.8cm}{\includegraphics[]{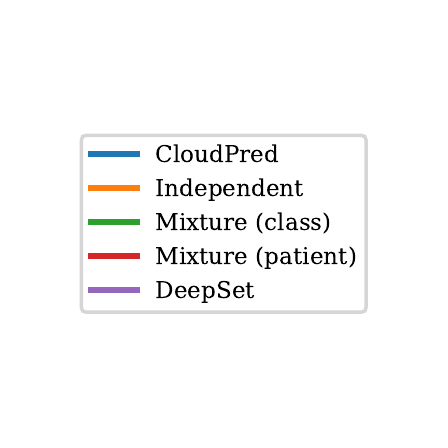}}
            \end{center}
        \end{subfigure}
        \begin{subfigure}[b]{0.64\textwidth}
            \begin{center}
                \includegraphics[]{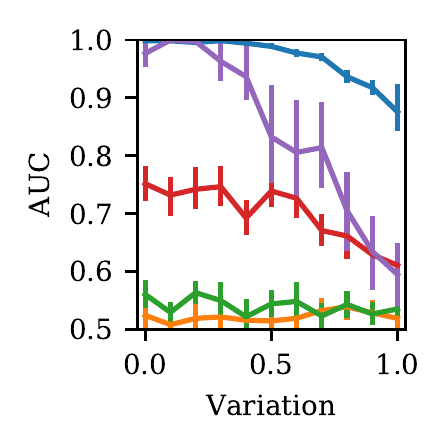}
                \includegraphics[]{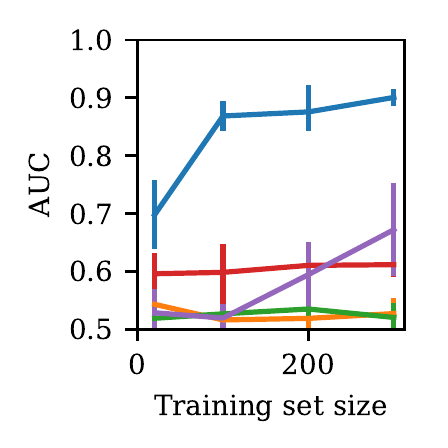}
                \caption{Increase or decrease of CD4 both indicate disease.}
                \label{fig: even}
            \end{center}
        \end{subfigure}
        \\
        \begin{subfigure}[b]{0.35\textwidth}
            \begin{center}
                \includegraphics[]{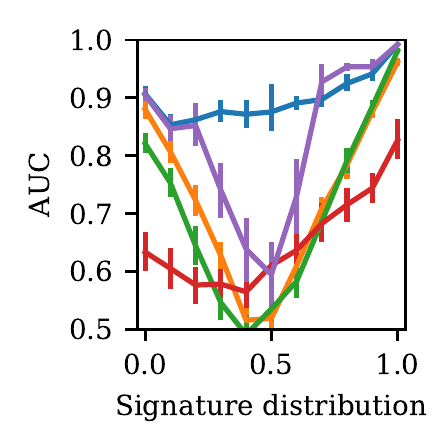}
                \caption{Varying signature prevalence.}
                \label{fig: two}
            \end{center}
        \end{subfigure}
        \begin{subfigure}[b]{0.64\textwidth}
            \begin{center}
                \includegraphics[]{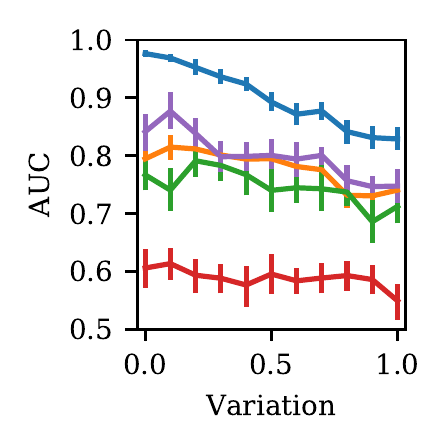}
                \includegraphics[]{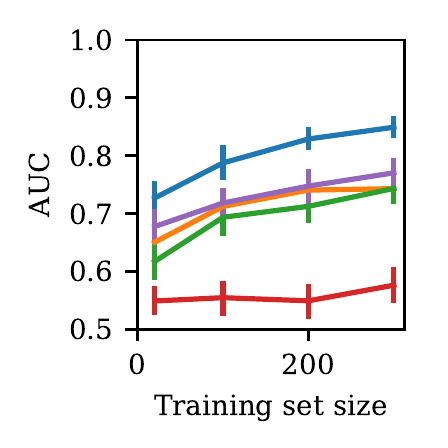}
                \caption{Interaction between cell types.}
                \label{fig: interact}
            \end{center}
        \end{subfigure}
        \caption{Comparison of CloudPred and alternative methods in a variety of simulated settings. (a) Both increases and decreases in CD4 T helper cells are ``disease signatures'' and are present in the dataset. Performance is studied for varying levels of patient-to-patient variation and number of training patients. (b) The frequency of the two disease signatures is varied from the case where only decreases are present (signature distribution 0) to the the case where only increases are present (signature distribution 1). (c) Both increases in the prevalence of CD4 T helper cells and CD14 Monocytes are required to indicate disease. Confidence intervals indicate bootstrapped 95\% confidence intervals.}
        \label{fig:simulations}
    \end{center}
\end{figure}
We first consider the challenge of identifying multiple signatures of a diseases.
We simulate both the increases and decreases in the prevalence of CD4 helper T cells as two possible signatures of disease (Fig.~\ref{fig: even}).
We first study the effect of patient-to-patient variability (Fig.~\ref{fig: example_variation}).
We simulate patient-to-patient variability by varying the prevalence of cell types unrelated to the disease signature.
It is easier to predict phenotype when the patient-to-patient variability is low---this corresponds to the a variation close to 0 in our experiments, where cell types unrelated to the disease have equal prevalences in all patients.
More realistic, and more challenging, settings correspond to a variation of 1, where we simulate the full variation reported by Brodin and Davis \cite{brodin2017nri}.
We find that all methods degrade in performance as the variation between patients is increased (Fig.~\ref{fig: even} left).
However, CloudPred remains robust even when the level of patient-to-patient variation is increased to realistic levels.
We then consider the performance as the number of training patients is changed with full patient-to-patient variation.
We find that even when the training set is reduced to 10 healthy controls and 10 diseased patients, CloudPred is still able to perform reasonably well (Fig.~\ref{fig: even} right).
All the methods improve in performance as the number of training patients is increased, but CloudPred reliably performs well across all cases.

Next, we consider the difficulty of identifying multiple signatures of a diseases as the prevalence of the two signatures is shifted in Fig.~\ref{fig: two}. A signature distribution of 0 represents the case where all diseased patients have decreases in frequency of CD4 T helper cells; a signature distribution of 1 represents the case where all diseased patients have increases in frequency CD 4 T helper; a signature distribution of 0.5 represents the case where the two signatures each appear in half of the disease patients.
We find that a mixture of the two cases is the most challenging case. In this scenario, CloudPred maintains high accuracy ($\mathrm{AUC} > 0.8$), while the comparison methods' AUC drop to less than 0.6. 

Finally, we investigate a setting where there are interactions between two cell types, where both cell types must increase in frequency to indicate a disease (Fig.~\ref{fig: example_interaction}).
In our simulations, both the prevalence of CD14 Monocytes and CD4 T helper cells must increase to be indicative of disease.
In this case, we find all of the baselines are able to perform better than chance, and CloudPred achieves the best accuracy in all of the simulations (Fig.~\ref{fig: interact}).

\subsection{Lupus and race classification using single-cell RNA-seq}

\begin{table}[thbp]
    \begin{center}
    \footnotesize
        \caption{AUC on the lupus scRNA-seq dataset (with 95\% bootstrapped confidence intervals)}
        \label{table: res}
        \begin{tabular}{lccc}
            \toprule
            Method & \multicolumn{3}{c}{Target} \\
            \cline{1-1} \cline{2-4}
            & Disease (AUC) & Race (AUC) & Monocyte (R2) \\
            \midrule
            CloudPred            & 0.98 (0.97-0.99) & 0.87 (0.85-0.89) & 0.97 (0.96-0.98) \\
            Independent          & 0.98 (0.97-0.99) & 0.74 (0.71-0.76) & 0.95 (0.94-0.95) \\
            Mixture (Class)   & 0.93 (0.91-0.95) & 0.76 (0.74-0.78) & ---               \\
            Mixture (Patient) & 0.77 (0.74-0.80) & 0.64 (0.61-0.66) & ---               \\
            Deepset              & 0.98 (0.98-0.99) & 0.78 (0.75-0.81) & 0.86 (0.78-0.95) \\
            \bottomrule
        \end{tabular}
    \end{center}
\end{table}

The extensive simulations give us confidence that CloudPred can make accurate predictions on challenging data.
Next, we analyze a real, recently generated scRNA-seq dataset consisting of 142 patients consisting of 566,453 cells \cite{SLE_draft}.
From the expression profile in this dataset, we predict disease state (22 healthy, 120 lupus), race (80 European descent, 62 Asian descent), and monocyte composition estimated from the Complete Blood Count reported in the UCSF Electronic Health Record.
The results for all methods are reported in Fig.~\ref{table: res}.

There are 32,738 mRNA species counted, resulting in a 32,738 dimensional vector of non-negative integers.
The high dimension of the scRNA-seq data makes downstream analysis difficult to interpret, and the dimension is very large in comparison to the number of patients, resulting in a high risk of overfitting.
As a result, we project the RNA counts onto the top 100 principal components to reduce the dimensionality of the data \cite{hotelling1933analysis,hotelling1992relations}.

We found that CloudPred and several of the other methods perform well at predicting disease status. This could be because lupus has relatively substantial transcriptome signature
\cite{crow2014type,baechler2004emerging,obermoser2010interferon}.
In contrast, predicting race is a more difficult problem, causing all of the comparison methods to perform significantly worse than CloudPred.
We additionally apply CloudPred, the independent model, and Deepset to predict monocyte composition in the blood, which would typically require a separate lab test.
The mixture-based baselines depend on using discrete classes, and are unable to be applied to the continuous monocyte composition setting. CloudPred performs the best here as well. 

To ensure that our method and the baselines generalize across patients, we train on 50\% of patients and we hold out and 25\% for validation and 25\% for testing.
The training set is used to learn parameters, and the validation set is used to pick hyperparameters for all methods.
Due to the imbalanced nature of the dataset, we report the area under the receiver operating characteristics (AUROC).
All experiments are run for 25 random splits of the dataset, and uncertainties represent 95\% confidence intervals computed with 10,000 bootstrapped samples.

\begin{figure}[htb!]
    \begin{center}
        \begin{subfigure}[b]{0.25\textwidth}
            \begin{center}
                \includegraphics[scale=0.8]{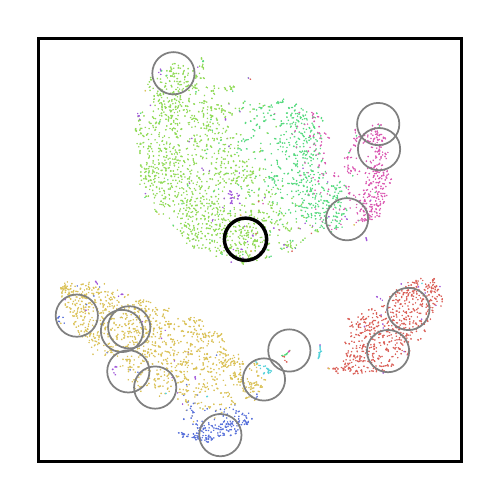}
                \caption{Lupus}
            \end{center}
        \end{subfigure}
        \begin{subfigure}[b]{0.25\textwidth}
            \begin{center}
                \includegraphics[scale=0.8]{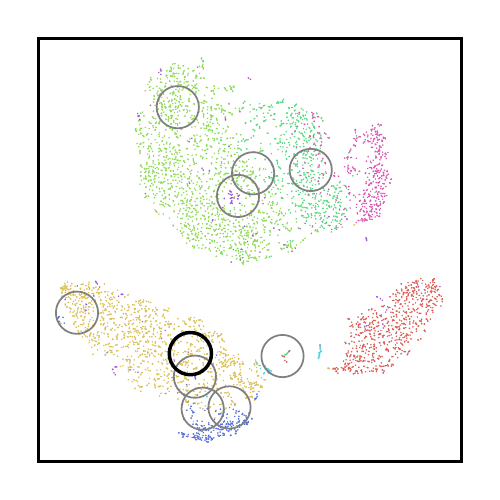}
                \caption{Race}
            \end{center}
        \end{subfigure}
        \begin{subfigure}[b]{0.25\textwidth}
            \begin{center}
                \includegraphics[scale=0.8]{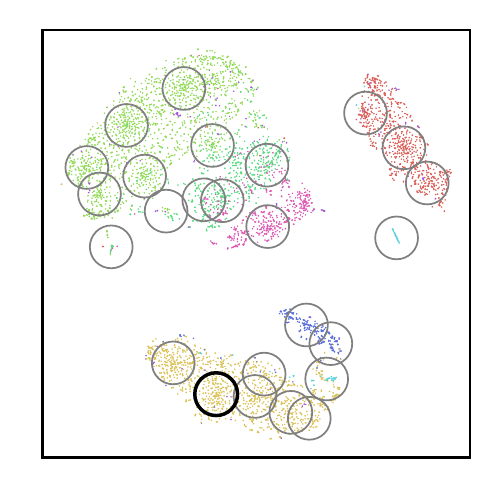}
                \caption{Monocyte}
            \end{center}
        \end{subfigure}
        \begin{subfigure}[b]{0.22\textwidth}
            \begin{center}
                \raisebox{0.5cm}{\includegraphics[scale=0.8]{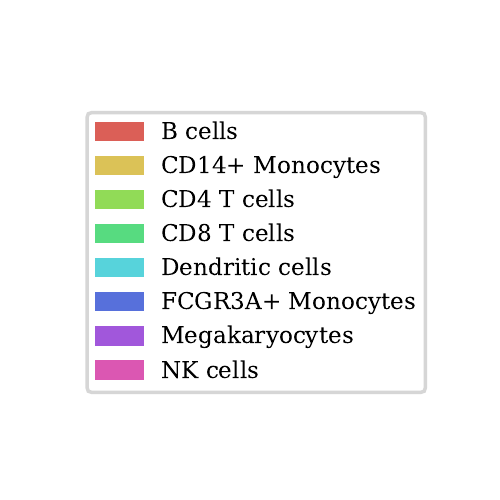}}
            \end{center}
        \end{subfigure}
        \caption{Lupus scRNA-seq dataset visualized with t-SNE \cite{maaten2008visualizing}. All of the 35 healthy and disease test patients are shown together here. Colors indicate expert annotated cell types. Circles are the the CloudPred learned cell subpopulations for predicting lupus status (a), race (b) and monocyte fraction (c). The single most predictive subpopulation is shown in black and additional clusters shown in gray.}
        \label{fig: lupus}
    \end{center}
\end{figure}

Because CloudPred learns an intuitive mixture model, it directly provides interpretations and insights into which cell populations are informative markers of each phenotype. For example, for predicting lupus, CloudPred learns that the prevalence of a single cluster of CD4 T cells is highly predictive (Fig.~\ref{fig: lupus}a). Using this the abundance of this cluster by itself achieves AUROC of 0.95. This subpopulation of cells is not specifically annotated \emph{a priori}, and its discovery is one of the contributions of CloudPred. Differential expression analysis identified that this lupus associated cluster of CD4 T cells is distinguished by have higher expression for several key immune and inflammatory markers S100A9, LYZ, S100A8, TYROBP, HLA-DRA, CD74, CST3, LGALS1, S100A4 and CTSS. 
Similarly, for predicting race, CloudPred identified a cluster of CD14+ monocytes as the most informative (Fig.~\ref{fig: lupus}b). Increasing prevalence of this subpopulation of cells is highly indicative of being Asian (AUROC 0.86).  

\paragraph{Sensitivity to number of patients and cells}

\begin{figure}[htb!]
    \begin{center}
    \begin{minipage}[c]{0.75\textwidth}
        \begin{subfigure}[b]{0.325\textwidth}
            \begin{center}
                \includegraphics[scale=0.8]{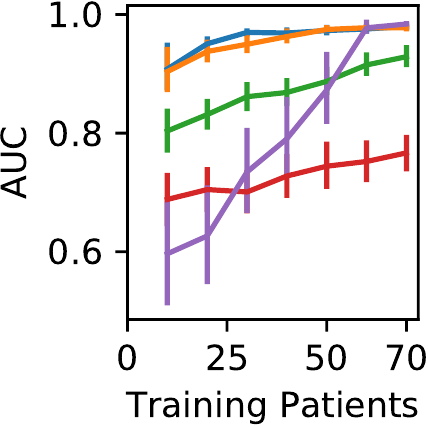}
                \caption{Lupus}
            \end{center}
        \end{subfigure}
        \begin{subfigure}[b]{0.325\textwidth}
            \begin{center}
                \includegraphics[scale=0.8]{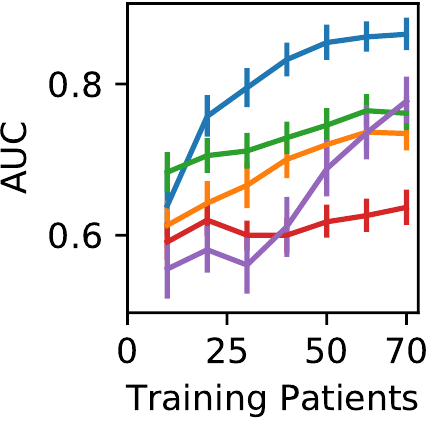}
                \caption{Race}
            \end{center}
        \end{subfigure}
        \begin{subfigure}[b]{0.325\textwidth}
            \begin{center}
                \includegraphics[scale=0.8]{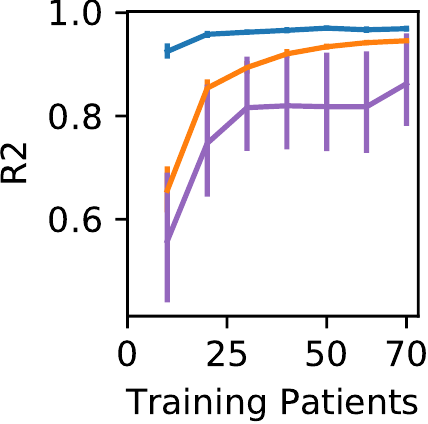}
                \caption{Monocyte}
            \end{center}
        \end{subfigure}
        \\
        \begin{subfigure}[b]{0.325\textwidth}
            \begin{center}
                \includegraphics[scale=0.8]{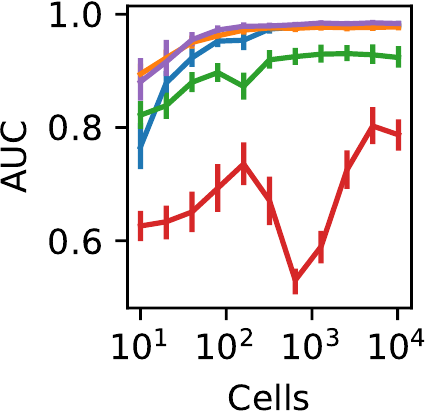}
                \caption{Lupus}
            \end{center}
        \end{subfigure}
        \begin{subfigure}[b]{0.325\textwidth}
            \begin{center}
                \includegraphics[scale=0.8]{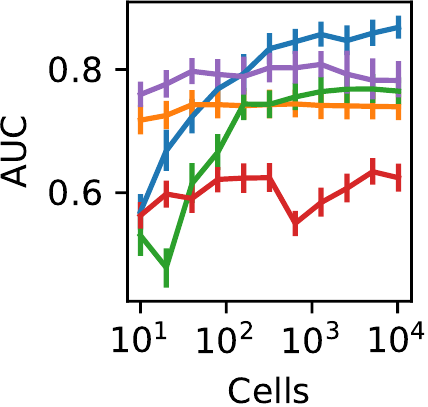}
                \caption{Race}
            \end{center}
        \end{subfigure}
        \begin{subfigure}[b]{0.325\textwidth}
            \begin{center}
                \includegraphics[scale=0.8]{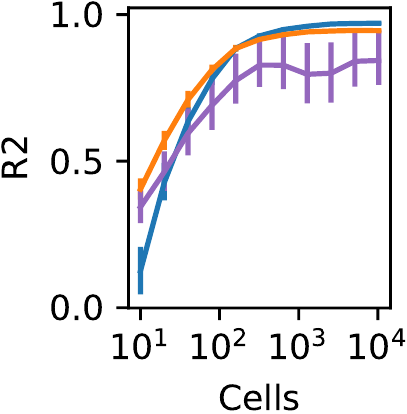}
                \caption{Monocyte}
            \end{center}
        \end{subfigure}
        \end{minipage}
        \begin{minipage}[c]{0.24\textwidth}
        \begin{subfigure}[b]{0.24\textwidth}
            \begin{center}
                \raisebox{-1.2cm}{\includegraphics[]{fig/legend.pdf}}
            \end{center}
        \end{subfigure}
        \end{minipage}
    \end{center}
    \caption{Comparison of CloudPred and alternative methods in different training settings using the Lupus scRNA-seq dataset. The number of training patients is varied for predicting lupus (a), race (b), and monocyte composition (c), and the number of cells is varied for predicting lupus (d), race (e), and monocyte composition (f).}
    \label{fig: hyperparam}
\end{figure}

We study the performance of all methods as the number of patients and cells are varied in the lupus dataset (Fig.~\ref{fig: hyperparam}).
The performance for the methods generally improve as the number of patients and cells increase.
CloudPred quickly performs well as the number of patient increases, especially compared to DeepSets, which is the most complex of the methods.
The performance for the methods are mostly stable after 1,000 cells are included, suggesting that current multiplexed sequencing techniques could be focused on sequencing more patients, rather than cells.

\paragraph{CloudPred scalability} CloudPred  efficiently scales to modern scRNA-seq datasets. Even on samples with 100,000 sequenced cells, CloudPred makes prediction in less than one second using a standard CPU.

\section{Discussion}

As single cell techniques become more ubiquitous, they open up exciting new opportunities for clinical prognosis and decision support. The ability to predict phenotypes from single cell data thus becomes increasingly critical. The heterogeneous nature of the single cell data poses challenges for standard computational biology approaches, which typically assume that every individual has similar number of measurements. In the case of single cells, each measurement corresponds to one cell and the number of cells could differ substantially across individuals. 

We developed and systematically evaluated several methods for predicting phenotype from scRNA-seq data. Across several experiments using both synthetic and real data, CloudPred achieves the best performance. CloudPred automatically infers from the data which subpopulations of cells are the most relevant for predicting the phenotype and works well even when trained on only 20 patients. Moreover it directly generates hypothesis about salient cell types.

CloudPred leverages the same type of end-to-end differentiable optimization that is the engine behind recent deep learning advances. However, CloudPred deploys this differentiable learning within the framework of a biologically motivated mixture model. This enables the algorithm to be more efficient than typical machine learning approaches, which requires large numbers of training examples. This also distinguishes CloudPred from unsupervised clustering and mixture modeling of single cells, where the clusters are learned from the scRNA-seq without leveraging the patient phenotype. CloudPred initializes with an unsupervised mixture model but then uses the phenotype to adjust and optimize each mixture component.
CloudPred is also directly interpretable. For example, CloudPred identified a particular subpopulation within CD4 T cells as the most informative for predicting lupus vs. healthy. Previous works have demonstrated the important role of CD4 T helper cells in lupus and other autoimmune diseases. An interesting direction of future work is to further investigate the functions of the specific CD4 T cells identified by CloudPred.   

Our analysis also highlight the value of systematic simulations of patient single cell data. We provide a framework to generate synthetic patient scRNA-seq data, from sorted blood cells, which captures patient variability and potential interactions between disease signatures. Such synthetic data is especially useful for single cell prediction analysis since the ground truth mechanism are often unknown.

We have released the software for both CloudPred and for generating synthetic data which could be broadly useful for the single cell community
Using scRNA-seq data in clinical settings still has challenges to overcome, including the use of data from multiple experiments, which may be sequenced using different protocols and have distinct batch effects, and predicting more fine-grained disease subtypes.
We hope our release of CloudPred can advance further work in predicting disease status from scRNA-seq data, such as improving upsupervised preprocessing of scRNA-seq data and improving models for predicting phenotypes.

\paragraph{Code availability}
An Pytorch based implementation of CloudPred, along with scripts to run experiments are available at \url{https://github.com/bryanhe/CloudPred}.

\paragraph{Data availability}
The datasets used in this paper are available from the original studies where the data was generated.

\bibliographystyle{ws-procs11x85}
\bibliography{bibliography}

\begin{thebibliography}{10}

\bibitem{tang2011nm}
F.~Tang, K.~Lao and M.~A. Surani, Development and applications of single-cell
  transcriptome analysis, {\em Nature methods} {\bf 8}  (2011).

\bibitem{eberwine2014promise}
J.~Eberwine, J.~Sul, T.~Bartfai and J.~Kim, The promise of single-cell
  sequencing, {\em Nature methods} {\bf 11}, p.~25  (2014).

\bibitem{zou2019primer}
J.~Zou, M.~Huss, A.~Abid, P.~Mohammadi, A.~Torkamani and A.~Telenti, A primer
  on deep learning in genomics, {\em Nature genetics} {\bf 51}, 12  (2019).

\bibitem{luecken2019current}
M.~D. Luecken and F.~J. Theis, Current best practices in single-cell {RNA}-seq
  analysis: a tutorial, {\em Molecular systems biology} {\bf 15}  (2019).

\bibitem{abid2018exploring}
A.~Abid, M.~J. Zhang, V.~K. Bagaria and J.~Zou, Exploring patterns enriched in
  a dataset with contrastive principal component analysis, {\em Nature
  communications} {\bf 9}, p. 2134  (2018).

\bibitem{lazebnik2008supervised}
S.~Lazebnik and M.~Raginsky, Supervised learning of quantizer codebooks by
  information loss minimization, {\em IEEE transactions on pattern analysis and
  machine intelligence} {\bf 31}, 1294  (2008).

\bibitem{chidester2018discriminative}
B.~Chidester, M.~N. Do and J.~Ma, Discriminative bag-of-cells for
  imaging-genomics, in {\em Proceedings of the Pacific Symposium\/},  (PSB,
  2018).

\bibitem{zhang2019valid}
J.~M. Zhang, G.~M. Kamath and N.~T. David, Valid post-clustering differential
  analysis for single-cell {RNA}-seq, {\em Cell systems} {\bf 9}, 383  (2019).

\bibitem{scikit-learn}
F.~Pedregosa, G.~Varoquaux, A.~Gramfort, V.~Michel, B.~Thirion, O.~Grisel,
  M.~Blondel, P.~Prettenhofer, R.~Weiss, V.~Dubourg, J.~Vanderplas, A.~Passos,
  D.~Cournapeau, M.~Brucher, M.~Perrot and E.~Duchesnay, Scikit-learn: Machine
  learning in {P}ython, {\em Journal of Machine Learning Research} {\bf 12},
  2825  (2011).

\bibitem{paszke2017automatic}
A.~Paszke, S.~Gross, F.~Massa, A.~Lerer, J.~Bradbury, G.~Chanan, T.~Killeen,
  Z.~Lin, N.~Gimelshein, L.~Antiga {\em et~al.}, Pytorch: An imperative style,
  high-performance deep learning library, in {\em Advances in neural
  information processing systems\/},  (NeurIPS, 2019).

\bibitem{chen2018biorxiv}
S.~Chen, P.~Rivaud, J.~H. Park, T.~Tsou, E.~Charles, J.~R. Haliburton,
  F.~Pichiorri and M.~Thomson, Dissecting heterogeneous cell populations across
  drug and disease conditions with popalign, {\em Proceedings of the National
  Academy of Sciences} {\bf 117}, 28784  (2020).

\bibitem{zaheer2017nips}
M.~Zaheer, S.~Kottur, S.~Ravanbakhsh, B.~Poczos, R.~R. Salakhutdinov and A.~J.
  Smola, Deep sets, in {\em Advances in Neural Information Processing
  Systems\/},  (NeurIPS, 2017).

\bibitem{qi2017cvpr}
C.~R. Qi, H.~Su, K.~Mo and L.~J. Guibas, Pointnet: Deep learning on point sets
  for 3d classification and segmentation, in {\em Computer Vision and Pattern
  Recognition\/},  (2) (IEEE, 2017).

\bibitem{ghorbani2019interpretation}
A.~Ghorbani, A.~Abid and J.~Zou, Interpretation of neural networks is fragile,
  in {\em AAAI Conference on Artificial Intelligence\/},  (AAAI, 2019).

\bibitem{brodin2017nri}
P.~Brodin and M.~M. Davis, Human immune system variation, {\em Nature reviews
  immunology} {\bf 17}, p.~21  (2017).

\bibitem{brodin2015cell}
P.~Brodin, V.~Jojic, T.~Gao, S.~Bhattacharya, C.~J.~L. Angel, D.~Furman,
  S.~Shen-Orr, C.~L. Dekker, G.~E. Swan, A.~J. Butte {\em et~al.}, Variation in
  the human immune system is largely driven by non-heritable influences, {\em
  Cell} {\bf 160}, 37  (2015).

\bibitem{shen2016cell}
S.~S. Shen-Orr, D.~Furman, B.~A. Kidd, F.~Hadad, P.~Lovelace, Y.~Huang,
  Y.~Rosenberg-Hasson, S.~Mackey, F.~A.~G. Grisar, Y.~Pickman {\em et~al.},
  Defective signaling in the {JAK-STAT} pathway tracks with chronic
  inflammation and cardiovascular risk in aging humans, {\em Cell systems} {\bf
  3}, 374  (2016).

\bibitem{efron1994introduction}
B.~Efron and R.~J. Tibshirani, {\em An introduction to the bootstrap} (CRC
  press, 1994).

\bibitem{SLE_draft}
M.~Subramaniam, Implementing and applying multiplexed single cell
  {RNA}-sequencing to reveal context-specific effects in systemic lupus
  erythematosus, PhD thesis, UC San Francisco, (UCSF, \ 2019).

\bibitem{hotelling1933analysis}
H.~Hotelling, Analysis of a complex of statistical variables into principal
  components., {\em Journal of educational psychology} {\bf 24}, p. 417
  (1933).

\bibitem{hotelling1992relations}
H.~Harold, Relations between two sets of variates, {\em Biometrika} {\bf 28},
  321  (1936).

\bibitem{crow2014type}
M.~K. Crow, Type {I} interferon in the pathogenesis of lupus, {\em The Journal
  of Immunology} {\bf 192}, 5459  (2014).

\bibitem{baechler2004emerging}
E.~C. Baechler, P.~K. Gregersen and T.~W. Behrens, The emerging role of
  interferon in human systemic lupus erythematosus, {\em Current opinion in
  immunology} {\bf 16}, 801  (2004).

\bibitem{obermoser2010interferon}
G.~Obermoser and V.~Pascual, The interferon-$\alpha$ signature of systemic
  lupus erythematosus, {\em Lupus} {\bf 19}, 1012  (2010).

\bibitem{maaten2008visualizing}
L.~van~der Maaten and G.~Hinton, Visualizing data using t-{SNE}, {\em Journal
  of machine learning research} {\bf 9}, 2579  (2008).

\end{thebibliography}

\end{document}